\documentclass[twocolumn,amsmath,amssymb]{revtex4}

\pdfoutput=1

\usepackage{graphicx}% Include figure files
\usepackage{dcolumn}% Align table columns on decimal point
\usepackage{bm}% bold math
\newcommand{\etal}{\emph{et al.}}
\newcommand{\be}{\begin{equation}}
\newcommand{\ee}{\end{equation}}
\newcommand{\bfig}{\begin{figure}}
\newcommand{\efig}{\end{figure}}

\begin{document}
\title{Heat capacity peak at the quantum critical point of the transverse Ising magnet CoNb$_2$O$_6$.
}
\author{
{Tian Liang$^1$, S.M. Koohpayeh$^2$, J. W. Krizan$^3$, T. M. McQueen$^{2,4}$, R. J. Cava$^3$ and N. P. Ong$^{1,*}$}
}
\affiliation{
\mbox{Departments of Physics$^1$ and Chemistry$^3$, Princeton University, Princeton, NJ 08544}\\
{Institute for Quantum Matter$^2$, Department of Physics and Astronomy$^2$ and Department of Chemistry$^4$, Johns Hopkins University, Baltimore, MD 21218}\\
{$^*$Correspondence to: npo@princeton.edu}
}

\date{\today}
%\pacs{}
\begin{abstract} 
The transverse Ising magnet Hamiltonian describing the Ising chain in a transverse magnetic field is the archetypal example of a system that undergoes a transition at a quantum critical point (QCP). The columbite CoNb$_2$O$_6$ is the closest realization of the transverse Ising magnet found to date. At low temperatures, neutron diffraction has observed a set of discrete collective spin modes near the QCP. We ask if there are low-lying spin excitations distinct from these relatively high energy modes. Using the heat capacity, we show that a significant band of gapless spin excitations exists. At the QCP, their spin entropy rises to a prominent peak that accounts for 30$\%$ of the total spin degrees of freedom. In a narrow field interval below the QCP, the gapless excitations display a fermion-like, temperature-linear heat capacity below 1 K. These novel gapless modes are the main spin excitations participating in, and affected, by the quantum transition.  
\end{abstract}

%\pacs{}
%72.15.Rn, 03.65.Vf, 71.70.Ej, 73.25.+i

\maketitle                   % Produces the title
In the transverse Ising magnet (TIM), a magnetic field applied transverse 
to the easy axis of the spins induces
a zero-Kelvin phase transition from the magnetically ordered state to the disordered state.
Because it is the archetypal example of a system displaying quantum critical behavior~\cite{Sachdev}, the TIM is prominently investigated in many areas of topical interest, e.g. quantum magnetism~\cite{Aeppli,Coldea}, integrable field theories~\cite{E8,Delfino}, and investigations of novel topological excitations~\cite{Kitaev,Fendley,Alicea}. 
The columbite CoNb$_2$O$_6$ is the closest realization found to date of the TIM
in a real material. The spin excitations have been investigated by
neutron diffraction spectroscopy near the quantum critical point (QCP)~\cite{Coldea} and in the paramagnetic state~\cite{Cabrera}, THz spectroscopy~\cite{Armitage}, and $^{93}$Nb nuclear magnetic resonance~\cite{Imai}, but little is known about their thermodynamic properties at the QCP. Are there low-lying spin excitations distinct from the neutron-excited modes? What are their characteristics at the QCP?
We report the results of a detailed heat capacity experiment which address these questions.

\begin{figure*}[t]
\includegraphics[width=10 cm]{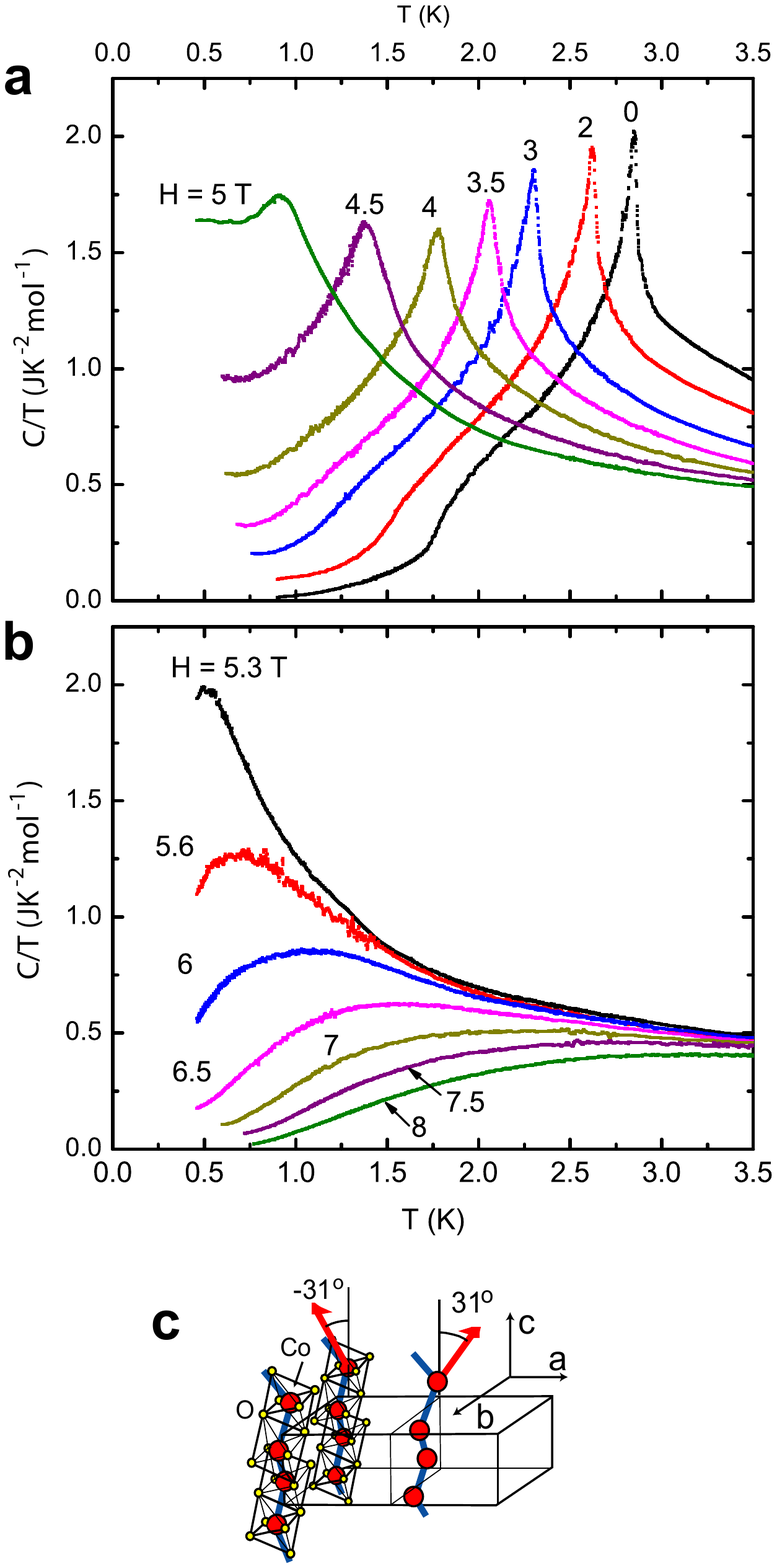}
\caption{\label{figCbT}  
{\bf The heat capacity of CoNb$_2$O$_6$ versus temperature measured in a transverse magnetic field.}
The field $\bf H$ is applied parallel to the $b$ axis. The quantity plotted is the heat capacity $C$ divided by temperature $T$.
Panel ({\bf a}): Curves of $C/T$ vs. $T$ at fixed $H <H_\mathrm{c}$ (= 5.24 T). In field $H$ = 0, a transition
to an incommesurate phase occurs at $T_\mathrm{c}(0)$ = 2.85 K. With increasing $H$, the transition $T_\mathrm{c}(H)$
is decreased. Below 1 K, $C/T$ approaches saturation instead of decreasing to 0.
At 5 T, $C/T$ is $T$ independent below 0.8 K. 
Panel ({\bf b}): Behavior of $C/T$ in the paramagnetic state ($H>H_\mathrm{c}$). Just above $H_\mathrm{c}$ (curve at $H$ = 5.3 T), $C/T$ falls 
montonically as $T$ increases above 0.5 K. For $H$ slightly above 5.6 T, a field-dependent gap $\Delta$ appears.
Panel ({\bf c}) shows the crystal structure of CoNb$_2$O$_6$. The easy axis (red arrows) is in the $a$-$c$ plane 
at an angle $\pm\,31^{\rm o}$ to the $c$-axis.
}
\end{figure*}

In CoNb$_2$O$_6$, the stacking of edge-sharing CoO$_6$ octahedra along the $c$-axis 
defines the Ising chain (inset in Fig. \ref{figCbT}b). The isolated chain is described by the TIM
Hamiltonian
\be
H_{1D} = -J_0\sum_n S^x_nS^x_{n+1} - \Gamma\sum_n S^z_n,
\label{eq:H1}
\ee 
with $J_0$ the ferromagnetic exchange along the easy axis $\bf c||x$ and $\Gamma$ the transverse field. $S^x_n$ and $S^z_n$ are, respectively, the $x$ and $z$ components of the spin operator at lattice site $n$.
In the $a$-$b$ plane, the chains assume a triangular coordination~\cite{Hanawa,Kobayashi,Balents}, with
antiferromagnetic interactions $|J_1|,\,|J_2|\ll J_0$ between adjacent chains. Geometric frustration effects lead to competing antiferromagetic and ferrimagnetic ground states~\cite{Balents}. In a magnetic field ${\bf H||b}$, CoNb$_2$O$_6$ exhibits a sharp transition to a 3D ordered phase 
at a critical temperature $T_\mathrm{c}(H)$ that decreases from 2.85 K (at $H=0$) to zero as $H\to H_\mathrm{c}$.

\vspace{1cm}\noindent{\bf \large Results}\\
\noindent {\bf Heat Capacity vs Temperature}
To investigate the low-energy spin excitations in CoNb$_2$O$_6$, we have measured its low-temperature heat capacity $C(T,H)$ by AC calorimetry over the $T$-$H$ plane [see Methods]. First, we discuss the curves of the heat capacity $C$ vs. $T$ measured in fixed $H$. Figure \ref{figCbT}a plots these curves as $C/T(T,H)$ vs. $T$ for $H<H_\mathrm{c}$ = 5.24 T.
In each curve, $C/T$ displays a prominent peak when $T$ crosses $T_\mathrm{c}(H)$. 
In zero $H$, $C/T$ decreases steeply below $T_\mathrm{c}(0)$, and approaches zero at 1 K, 
consistent with the existence of a full gap. The shoulder
feature near 1.7 K signals the transition from an incommensurate to commensurate AF phase~\cite{Kobayashi}. 
At finite $H$, we observe significant enhancement of
$C/T$ throughout the ordered phase. Instead of falling to zero, the curves become $T$ independent at low $T$
(curve at 5 T).
Between 4 and 5 T, the saturation value increases by more than a factor of 3. 
In the disordered phase ($H>H_\mathrm{c}$), we observe a profile that also reveals a gap $\Delta$,
but one that increases sharply with the reduced field $H - H_\mathrm{c}$ (Fig. \ref{figCbT}b).

\begin{figure*}[t]
\includegraphics[width=12 cm]{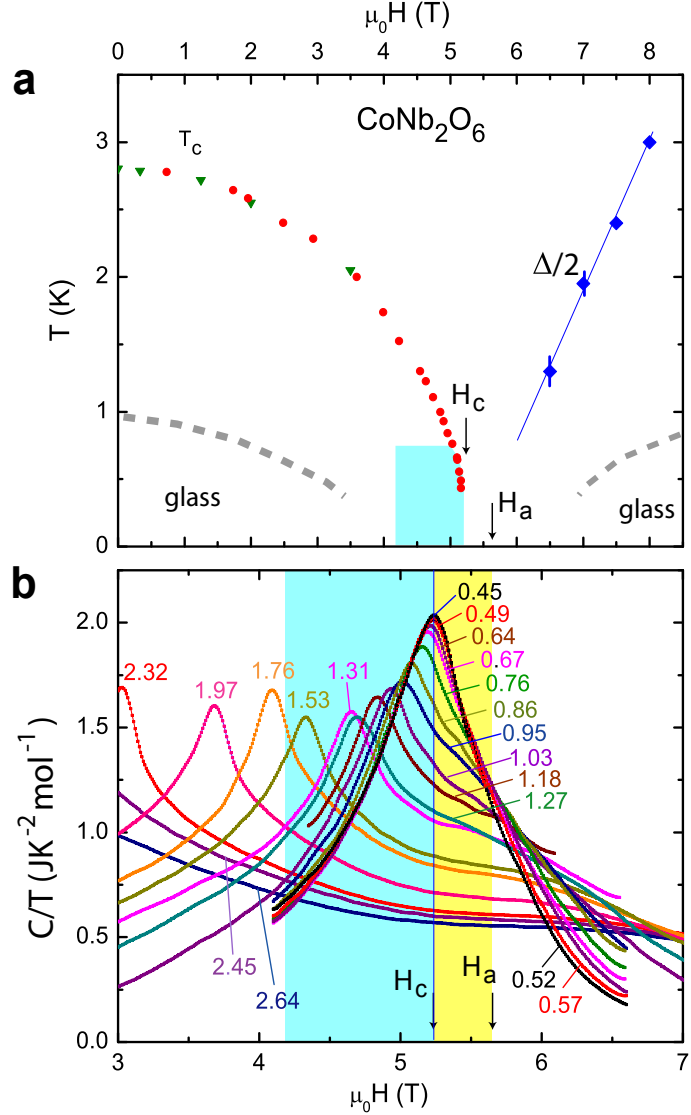}
\caption{\label{figphase}  
{\bf Phase diagram of CoNb$_2$O$_6$ in a transverse magnetic field and heat capacity peak at the QCP.}
Panel ({\bf a}): The phase diagram inferred from $C/T$. The transition $T_\mathrm{c}(H)$ defines the ordered phase (solid circles and triangles represent $T$-constant and $H$-constant measurements, respectively). Gapless excitations with $T$-independent $C/T$ 
are observed in the blue-shaded region ($4.2 T < H < H_\mathrm{c}$). The gap $\Delta$ above $H_a$ (solid diamonds) is inferred from fits to the 1D exact solution. The error bars are estimated from the goodness of the fits at each $H$. 
$H_a$ is the crossover field at which $d(C/T)/dT$ changes sign at low $T$. 
The dashed curves are nominal boundaries of the low-$T$ phases in which glassy behavior is observed.
Panel ({\bf b}): Curves of $C/T$ measured vs. $H$ at constant $T$ in Set 2. Above $\sim$1 K, $C/T$ climbs to a sharp peak when $H$ crosses the boundary $T_\mathrm{c}(H)$, and then falls monotonically. Below 1 K, however, the constant-$T$ contours lock to the left branch of the critical peak profile $(C/T)_0$ measured at 0.45 K (which peaks at $H_\mathrm{c}$ = 5.24 T). 
The locking implies $C/T$ is $T$-independent below 0.8 K. In both panels, the blue shaded regions represent the field interval within which the locking is observed.
Above $H_\mathrm{c}$, the contours are well-separated at all $T$. The derivative $d(C/T)/dT$ 
at low $T$ is negative for $H_\mathrm{c}<H<H_\mathrm{a}$ (yellow region), but positive for $H>H_a$.
}
\end{figure*}

\vspace{0.3cm}\noindent {\bf Heat Capacity vs Field}
To supplement the constant-$H$ curves in Fig. \ref{figCbT}, we performed measurements of $C/T$ vs. $H$ at constant $T$. Figure \ref{figphase}a shows the phase diagram obtained from combining the constant-$H$ and constant-$T$ curves. The boundary of the ordered phase, $T_\mathrm{c}(H)$ defined by the sharp peak in $C/T$, falls to zero as $H\to H_\mathrm{c}$ (solid circles and triangles). Above $H_\mathrm{c}$, the gap $\Delta$ in the disordered phase (solid diamonds) is estimated from fits to the free-fermion solution discussed below. The dashed curves are the nominal boundaries below which glassy behavior is observed (see below).

Figure \ref{figphase}b displays the constant-$T$ scans in the region close to the QCP. If the temperature is fixed at a relatively high value, e.g. $T$ = 1.76 K, $C/T(T,H)$ initially rises to a sharp peak as $H$ is increased from 0 to 4.1 T. Above 4.1 T, $C/T$ falls monotonically with no discernible feature at $H_\mathrm{c}$. As we lower $T$, the peak field shifts towards $H_\mathrm{c}$= 5.24 T, tracking $T_\mathrm{c}(H)$ in the
phase diagram in Panel (a). Below ~1.5 K, the constant-$T$ contours converge towards the prominent
profile measured at 0.45 K, which is our closest approximation to the critical peak profile $(C/T)_0$ at $T=0$. The contours under this profile reveal a remarkable structure. On the low-field side of the peak ($4.2\, {\rm T}< H < H_\mathrm{c}$), shaded blue in Figs. \ref{figphase}A and B, the contours lock to the critical peak profile as $T$ decreases. This implies that, if $H$ is fixed inside this interval, $C/T$ assumes the $T$-independent
value $(C/T)_0$ at low $T$. Hence the $T$-independent plateau seen in the curve at 5 T in Fig. \ref{figCbT}a is now seen to extend over the entire blue region. When $H$ exceeds $H_\mathrm{c}$, however, the locking pattern vanishes. The different $T$ dependencies reflect the distinct nature of the excitations on either side of $H_\mathrm{c}$. (Slightly above $H_\mathrm{c}$, the derivative $d(C/T)/dT$ changes from negative to positive at a crossover field $H_a\sim$ 5.6 T.)

\vspace{0.3cm}\noindent {\bf Spectrum of $C_{eff}$ and glassy response} In AC calorimetry, the spectrum of the effective (observed) heat capacity $C_{eff}(\omega)$ varies in a characteristic way with the measurement frequency $\omega$. For each representative local region of the $T$-$H$ phase diagram investigated, we measured $C_{eff}(\omega) \equiv P_0/(2\omega|\hat{T}_{ac}(\omega)|)$ over the frequency range 0.02 to 100 Hz, where $P_0$ is the applied power and $\hat{T}_{ac}$ is the complex temperature (see Methods). The spectrum of $C_{eff}(\omega)$ has a hull-shaped profile characterized by the two characteristic times $\tau_1$ (set by the sample's parameters) and $\tau_{ext}$ set by coupling to the bath (defined in Methods). In both the low-$\omega$ and high-$\omega$ regions ($\omega\ll 1/\tau_{ext}$ and $\omega\gg 1/\tau_1$, respectively) $C_{eff}(\omega)$ rises steeply above the true (equilibrium) heat capacity $C$. However, there exists a broad frequency range in-between where $C_{eff}(\omega)$ is nearly $\omega$-independent and equal to the intrinsic equilibrium heat capacity $C$ of the sample. All results reported here are taken with $\omega$ within this sweet spot. Within the regions denoted as glassy in Fig. \ref{figphase}, the spectrum is anomalous (Methods). Instead of the hull-shaped spectrum, the measured $C_{eff}$ decreases monotonically over the accessible frequency range. We define these regions of the phase diagram as glassy. We note that the QCP region lies well away from the glassy regions.

\begin{figure*}[t]
\includegraphics[width=14 cm]{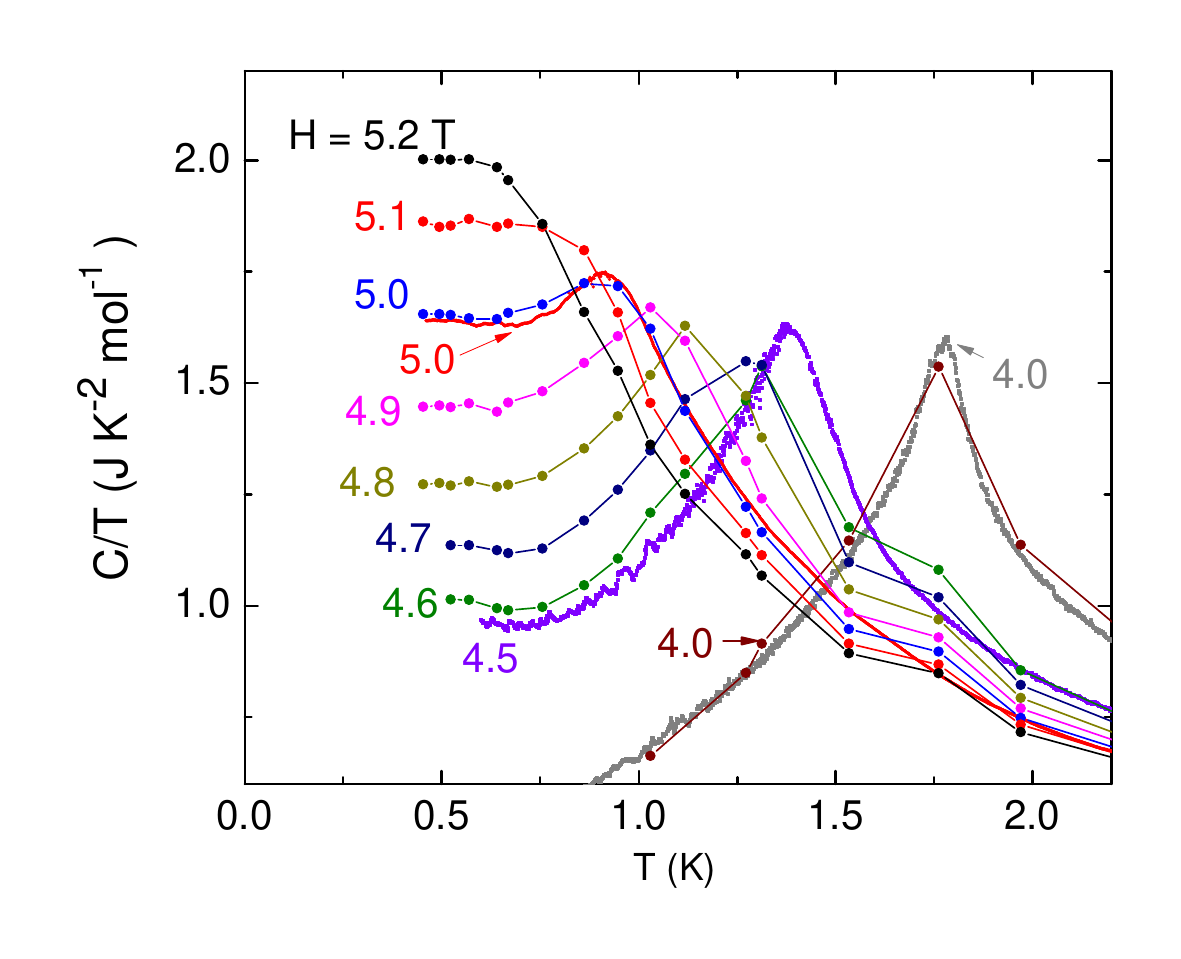}
\caption{\label{figexpand}  
{\bf Low temperature heat capacity and spin entropy in expanded scale near the critical field.}
The discrete symbols (solid circles) are values of $C/T$ extracted from continuous measurements of $C/T$ vs. $H$ 
at fixed $T$ (the constant-$T$ scans plotted in the phase diagram). Here they are plotted versus $T$ at fixed $H$ to bring out the fixed-field contours. Below 0.8 K, the values of $C/T$ saturate to a $T$-independent value that depends on $H$. These plateau values occur within the region of the phase diagram where the contours display locking behavior. To supplement the discrete data points, we also measured $C/T$ continuously versus $T$ at fixed $H$. These curves are shown as nearly continuous curves at $H$ = 4.0, 4.5 and 5.0 T. The agreement between the two distinct experiments is very close. 
}
\end{figure*}

\vspace{0.3cm}\noindent {\bf $T$-linear Heat Capacity at Critical Field} To make explicit the $T$-independent behavior below $H_\mathrm{c}$, we have extracted the values of $C/T(T,H)$ and replotted them in Fig. \ref{figexpand} as constant-$H$ curves for 8 values of $H$ between 4 and 5.2 T. As is evident, the curves approach a constant value when $T$ decreases below 0.8 K. The flat profiles reflect the locking of the contours described above. We have also plotted the constant-$H$ curves measured at 4, 4.5 and 5 T (continuous curves) to show the close agreement between the 2 sets of data. 
The critical peak profile in Fig. \ref{figphase}b and the $T$-independent contours shown in Fig. \ref{figexpand} are our key findings in this report.

The $T$-linear behavior of $C$ at low $T$ illuminates the nature of the low-lying excitations.
For fermions, $C$ is linear in $T$ and given by $C_F = (\pi^2/3)k_B^2T {\cal D}_F$,
where ${\cal D}_F$ is the density of states at the Fermi level. The Hamiltonian Eq. \ref{eq:H1} can be diagonalized by transforming to free fermions~\cite{Lieb,Lieb2} (see Free-fermion solution in Methods). The heat capacity of the isolated Ising chain may then be calculated~\cite{Pfeuty}. The issue whether the free fermions are artifacts or real observables is currently debated~\cite{Kitaev,Fendley}. However, our system is 3D with finite $J_1$ and $J_2$. To our knowledge, fermionic excitations in the ordered phase have not been anticipated theoretically.  

\vspace{0.3cm}\noindent {\bf Low Temperature Spin Entropy at Critical Field}
We next show that the critical peak profile ($(C/T)_0$ in Fig. \ref{figphase}b) accounts for a surprisingly 
large fraction of the total spin degrees of freedom (d.o.f.). Before extracting the spin entropy from the measured $C$, we need to subtract the phonon contribution.
Fortunately, the spin contributions may be readily distinguished from the phonon term. Following the procedure of
Hanawa \etal~\cite{Hanawa}, we have carried out this subtraction to isolate the spin part of the heat capacity $C_s(T,H)$ (see Methods).
The spin entropy is then given by the integral $S_s(T) = \int^T_0 dT' C_s(T')/T'$. 

\begin{figure*}
    \includegraphics[width=14cm]{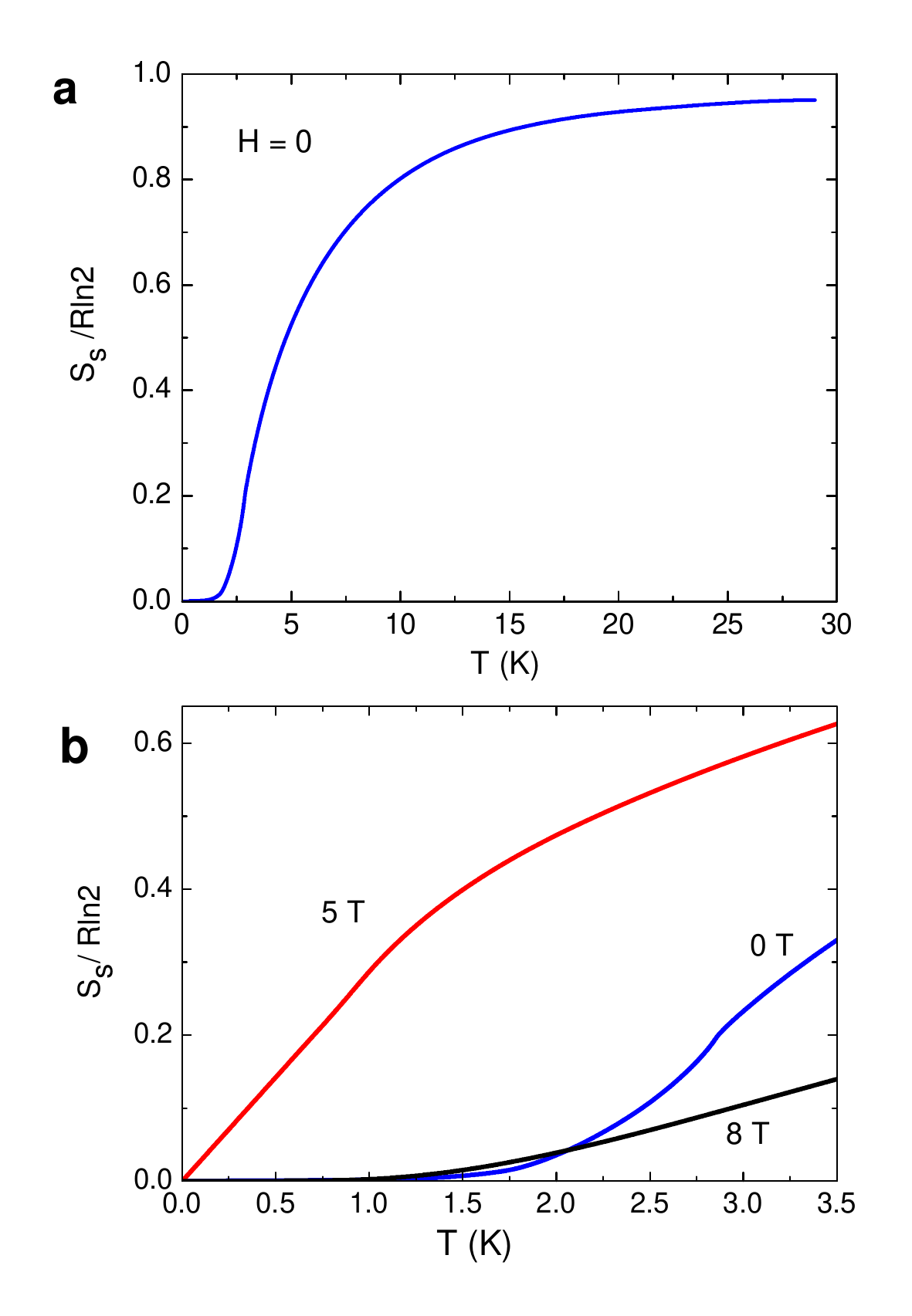}
    \caption{\label{Entropy} 
		{\bf The spin entropy of CoNb$_2$O$_6$ derived from its heat capacity.} To obtain the spin entropy $S_s$, we first subtract the phonon contribution from the measured heat capacity to isolate the spin contribution $C_s$. The spin entropy $S_s(T)$ is then obtained by integrating $C_s(T)/T$ with respect to $T$. Panel ({\bf a}) shows the profile of $S_s$ in zero magnetic field. At 20 K, 90$\%$ of the spin entropy frozen out at low $T$ is recovered. Although our measurements extend only to 30 K, the curve of $S_s$ is expected to asymptote to 1 above room temperature. Panel ({\bf b}) shows the spin entropy $S_s$ vs. $T$ at $H$ = 0, 5 and 8 T. $S_s$ is obtained by integrating the curves of $C_s/T$ after subtracting the phonon contribution (derived from the results of Hanawa \etal~on the nonmagnetic analog ZnNb$_2$O$_6$). At 5 T, $S_s$ accounts for nearly $\frac13$ of the spin d.o.f. at 1 K.
		} 
\end{figure*}

First, we verified that, at $H$ = 0, the curve of $S_s(T)$ obtained by integrating $C_s/T$ rises rapidly above 20 K to closely
approach the value $R\ln 2$ ($R$ is the universal gas constant), thus accounting for the total spin d.o.f. 
By integrating $C_s/T$ with respect to $T$, we obtain the total spin entropy $S_s(T)$. 
In Fig~\ref{Entropy}a, the variation of $S_s(T)$ vs. $T$ inferred from the data at zero $H$ is
plotted. Above $\sim$5 K, $S_s$ rises rapidly attaining 90$\%$ of $R\ln2$ by 20 K.

Our interest here is the behavior of $S_s$ at low $T$, which we plot in Fig. \ref{Entropy}b. 
In contrast to $S_s(T)$ at $H$ = 0 and 8 T, the curve for $S_s(T)$ at 5 T is strongly enhanced and 
varies linearly with $T$ with a slope equal to $C_s/T$. 
As $H\to H_\mathrm{c}$, the spin entropy rises to $\sim 30\%$ of $R\ln 2$ at 1 K. Hence, the gapless
excitations account for nearly $\frac13$ of the total spin d.o.f.

\vspace{0.3cm}\noindent {\bf Paramagnetic State Heat Capacity and the Free-Fermion Solution}
In the paramagnetic state above $H_\mathrm{c}$, it is instructive to compare the measured $C_s/T$ with the heat capacity
calculated from the free energy in the free-fermion solution~\cite{Pfeuty} (see Methods) given by
\be
C_s(T) = R\int^{\pi}_0 \frac{dk}{\pi} \left(\frac{\beta\varepsilon_k}{2}\right)^2\; {\rm sech}^2\left(\frac{\beta\varepsilon_k}{2}\right).
\label{eq:Cs}
\ee
where $R$ is the gas constant, $\beta = 1/(k_BT)$, and the energy of the fermions is 
$\varepsilon_k = \Gamma\sqrt{1+\lambda^2 + 2\lambda\cos k}$, with $ \lambda = J_0/2\Gamma$. For each field $H>H_\mathrm{c}$, we took $\Gamma$ and $\lambda$ as
adjustable parameters ($J_0$ is fixed at 21.4 K for all $H$).

\begin{figure*}
\includegraphics[width=14 cm]{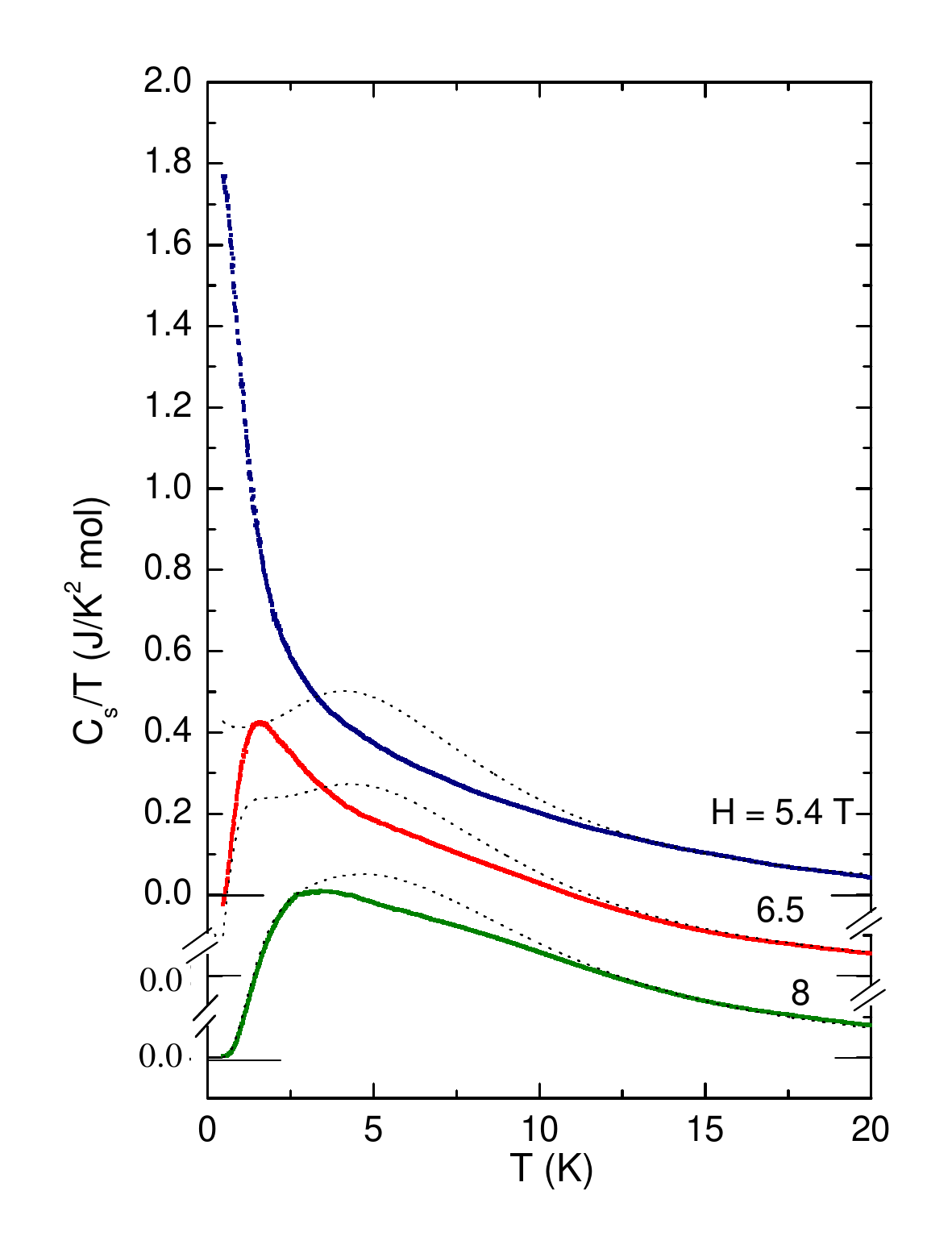}
\caption{\label{figfits}  
{\bf The heat capacity in the paramagnetic phase at 3 values of magnetic field above the critical value.} The phonon
contribution to the measured $C$ has been subtracted. For clarity, the curves have been
displaced vertically. The dashed curves are fits to the free-fermion solution using the values
$\Gamma$ = 11, 13.24 and 16.3 K for the curves at $H$ = 5.4, 6.5 and 8 T, respectively.
The corresponding values of $\lambda$ are 0.97, 0.806 and 0.655. 
The value of $J_0$ is fixed at 21.4 K. 
Deviations from the fits become pronounced as $H\to H_\mathrm{c}$.
}
\end{figure*}

As shown in Fig. \ref{figfits}, the fits (dashed curve) are reasonable 
only above 10 K. Below 10 K, deviations become increasingly prominent as we lower
$H$ towards $H_\mathrm{c}$. In particular, the striking divergence of the curve at $H$ = 5.4 T (as $T\to 0$) lies well beyond the reach of Eq. \ref{eq:Cs}. These deviations reveal that the incipient magnetic ordering effects extend deep into the paramagnetic phase. The curves in Figs. \ref{figphase} and \ref{figexpand} reveal how these deviations smoothly evolve into the gapless excitations. Models that include interchain exchange terms (e.g. as proposed in Ref. \cite{Cabrera}) are more realistic. Comparison of our data with the behavior of the $C_s$ predicted near the QCP should be highly instructive. To our knowledge, solutions in the quantum regime have not been reported.

\vspace{0.3cm}\noindent {\bf Negligible Contribution of Nuclear Spin Degrees}
We discuss whether contributions of the nuclear spins to the heat capacity play any role in the experiment. The nuclear spins contribute as a Schottky term given by~\cite{Collan1} $C_N = Nk_B X^2\mathrm{e}^X/(\mathrm{e}^X + 1)^2$ (for the 2-level case), where $X = \Delta E/k_BT$, and $\Delta E$ is the energy splitting of the levels. $C_N$ peaks near $\Delta E/k_B$ (typically 10-30 mK~\cite{Collan1,Collan2}) and falls off as $(\Delta E/k_BT)^2\;\equiv \; A/T^2$ for $T\gg \Delta E/k_B$. The most favorable situation is when $H$ increases $\Delta E$ by the Zeeman energy, viz. $\Delta E = \mu\mu_N H$, where $\mu$ is the nuclear moment and $\mu_N = 0.37$ mK/T the nuclear Bohr magneton. We have $\mu$ = 6.17 for $^{93}$Nb and 4.63 for $^{59}$Co. The larger moment gives $A$ = 0.032 mJK/mol at 5 Tesla. At $T$ = 1 K, this yields values for $C_N$ that are extremely small (by a factor of 10$^{5}$) compared with $C$ displayed in Figs. \ref{figCbT} and \ref{figexpand}. Hence, the nuclear spin d.o.f. cannot be resolved in our experiment.

\vspace{1cm}\noindent{\bf \large Discussion and Conclusions}\\
In the isolated Ising chain, the excitations are domain walls (kinks) which separate 
degenerate spin-$\uparrow$ from spin-$\downarrow$ domains. In our 3D system, the self-consistent
fields derived from $J_1$ and $J_2$ lift the degeneracy. As a result, kinks and antikinks interact via a linear potential (the energy cost of the unfavored domain) to form bound pairs~\cite{McCoy,Rutkevich}.
The quantized excitations of the bound pairs have been detected by neutron diffraction spectroscopy~\cite{Coldea} and by time-domain THz spectroscopy~\cite{Armitage} as discrete modes (the lowest mode has energy 1.2 meV at $H$ = 0 and 0.4 meV at 5 T). As $H\to H_\mathrm{c}$, the ratio of the two lowest modes approaches the golden ratio, consistent with the E8-Lie group spectrum~\cite{Coldea}. However, these modes are too high in energy to contribute to $C/T$ below 1 K. Rather, our experiment provides firm evidence for a band of low-lying, gapless spin excitations that are entirely distinct from the high-energy modes. The steep increase of the spin entropy $S_s$ at 5 T (Fig. \ref{Entropy}b) shows that $S_s$ has attained $30\%$ of its high-$T$ value already at 1 K. Thus the large anomalous peak centered at the QCP accounts for a significant fraction ($\sim\frac13$) of the total spin d.o.f. In addition to the remarkable spin modes observed at discrete energies by neutron diffraction spectroscopy and THz spectroscopy, a substantial fraction of the spin d.o.f. exists as (essentially) gapless modes which peak in weight at the QCP. How the two sets of excitations co-exist is a problem that confronts the theoretical description of the QCP in this material.

Perhaps the most surprising finding from the experiment is the $T$-independent profile of $C_s(T)/T$ below 1 K in the field interval $4.2<H<H_\mathrm{c}$ abutting $H_\mathrm{c}$. The results imply that the excitations obey Fermi-Dirac statistics. In the 1D TIM model, the solution obtained via the Jordan Wigner transformation (see Methods) yield free fermions. However, as mentioned above, it is uncertain whether these fermions are physically observable. Moreover, the interchain exchange in the real material~\cite{Cabrera} may render the free-fermion solutions inapplicable. The present finding that the $T$-linear behavior is confined to the QCP region where $C/T$ displays a prominent peak highlights serious gaps in our understanding of the QCP and the effects of strong quantum fluctuations in its vicinity. The heat capacity invites a detailed investigation of the quantum behavior at the QCP in realistic models applicable to CoNb$_2$O$_6$.

The heat capacity experiment shows that, in the vicinity of the QCP, the gapless modes constitute the dominant spin excitations that are affected by the quantum transition induced by the applied transverse $\bf H$. As seen in the set of curves in Fig. \ref{figphase}b, the QCP strongly affects $C/T$ vs. $H$ to produce a profile that, at 0.45 K, rises to a prominent peak at the critical field. We reason that the gapless modes are the relevant modes that participate in the quantum transition at $H_\mathrm{c}$. The dominant fluctuations associated with the quantum transition are inherent to these modes. From the spin entropy, we infer that they account for nearly $\frac13$ of the total spin degrees of freedom in the sample. As discussed, the gapless modes display a fermion-like heat capacity below 1 K over a broad region of the ordered phase below $H_\mathrm{c}$.

%%%%%%%%%%%%%%%%%%%%%%%%%%%%%%%%%%%%%%%%%%%
%%%%%%%%%%%%%%%%%%%%%%%%%%%%%%%%%%%%%%%%%%%
%%%%%%%%%%%%%%%%%%%%%%%%%%%%%%%%%%%%%%%%%%%
%%%%%%%%%%%%%%%%%%%%%%%%%%%%%%%%%%%%%%%%%%%

\vspace{1cm}\noindent{\bf \large Methods}\\
\noindent{\bf Crystal Growth}
The CoNb$_2$O$_6$ powder was packed and sealed into a rubber tube evacuated using a vacuum pump. The powder was then compacted into a rod, typically 6 mm in diameter and 70 mm long, using a hydraulic press under an isostatic pressure of 7$\times\,10^7$ Pa. After removal from the rubber tube, the rods were sintered in a box furnace at 1375 C for 8 h in air.

Single crystals of approximately 5 mm in diameter and 30 mm in length were grown from the feed rods in a four-mirror optical floating zone furnace (Crystal System Inc. FZ-T-4000-H-VII-VPO-PC) equipped with four 1-kW halogen lamps as the heating source. In all the growth processes, the molten zone was moved upwards with the seed crystal being at the bottom and the feed rod above it. Growths were carried out under 2 bar O$_2$-Ar (50/50) atmosphere with the flow rate of 50 mL/min, at the zoning rate of 2.5 mm/h, with rotation rates of 20 rpm for the growing crystal (lower shaft) and 10 rpm for the feed rod (upper shaft). In all runs, only one zone-pass was performed.

Phase identification and structural characterization were obtained using a Bruker D8 Focus X-ray diffractometer operating with Cu Kα radiation and Lynxeye silicon strip detector on finely ground powder from the crystal boules, while back-reflection X-ray Laue diffraction was utilized to check the crystalline qualities and orientations of the crystals. Measurements were carried out on oriented thin rectangular-shaped samples cut directly from the crystals using a diamond wheel.

\vspace{0.3cm}\noindent{\bf AC Calorimetry}
The heat capacity was measured using the AC calorimetry technique~\cite{Sullivan} on a crystal of CoNb$_2$O$_6$ (approximately 1 mm $\times$ 3 mm $\times$ 0.5 mm along the $a$, $b$ and $c$ axes, respectively) in a magnetic field $\bf H$ applied along the $b$-axis. 
Using an AC current of frequency $\frac12\omega$, we applied the AC power 
$(P_0/2)\exp(i\omega t)$ to the sample via a 1-k$\Omega $ RuO$_2$ thin-film resistor ($P_0$ ranged from $\sim 0.4\, \mu $W to 5 $\mu$W for measurements below 4 K, and $\sim 5\, \mu$W to $400\, \mu $W from 4 K to 30 K). The (complex) temperature $\hat{T}_{ac}\exp(i\omega t)$ was detected at the thermometer (a 20-k$\Omega $ RuO$_2$ thin-film resistor). In the slab geometry of Ref. \cite{Sullivan}, $\hat{T}_{ac}$ is given by
\be
\hat{T}_{ac}(\omega) =\frac{P_0/2K_b}{\cosh \hat{\theta} + (K_{int}\hat{\theta}/K_b)  \sinh\hat{\theta}},
\label{eq:Tac}
\ee
where $\hat{\theta}$ is the complex angle
\be
\hat{\theta} = \left[\frac{\omega C}{2K_{int}}\right]^{1/2} (1+i).
\label{eq:theta}
\ee
Here, $K_{int}$ is the thermal conductance of the sample, $K_b$ the thermal conductance between the sample and the thermal bath, and $C$ the heat capacity of the sample.

Taking into account that the thermal conductance between the thermometer and the sample, heater and the sample is finite, and under the condition that the internal relaxation time constant $\tau_1\equiv (\tau _{i}^2+\tau _{\theta}^2+\tau _h^2)^{1/2}$ (where $\tau _{i}=C/6K_{int}, \tau _{\theta}=C_{\theta}/K_{\theta}, \tau _h=C_h/K_h$, $C_{\theta}$ heat capacity of the thermometer, $C_h$ heat capacity of the heater, $K_{\theta}$ thermal conductance between the thermometer and the sample, $K_h$ thermal conductance between the heater and the sample) and the external relaxation time constant $\tau_{ext}\equiv C/K_b$ satisfy $\omega\tau_1\ll 1 \ll \omega\tau_{ext}$, Eq. \ref{eq:Tac} reduces to

\begin{eqnarray}
|\hat{T}_{ac}(\omega)| &=&\frac{P_0}{2\omega C} \left[1+(\omega \tau _1)^2+\frac{1}{(\omega \tau _{ext})^2}+\frac{2}{3}\frac{K_b}{K_{int}} \right]^{-1/2}
\label{eq:eff_heat}
\\
&\approx & \frac{P_0}{2\omega C}, \quad\quad (\omega\tau_1\ll 1 \ll \omega\tau_{ext}) 
\label{eq:sweet}
\end{eqnarray}
from which the heat capacity is obtained as 
$C=P_0/(2\omega|\hat{T}_{ac}(\omega^*)|\,)$, where $\omega^*$ is within the sweet spot ($\omega\tau_1\ll 1 \ll \omega\tau_{ext}$). The inequalities in Eq. \ref{eq:sweet} determine the optimal frequency $\omega^*$.

\begin{figure*}
    \includegraphics[width=14cm]{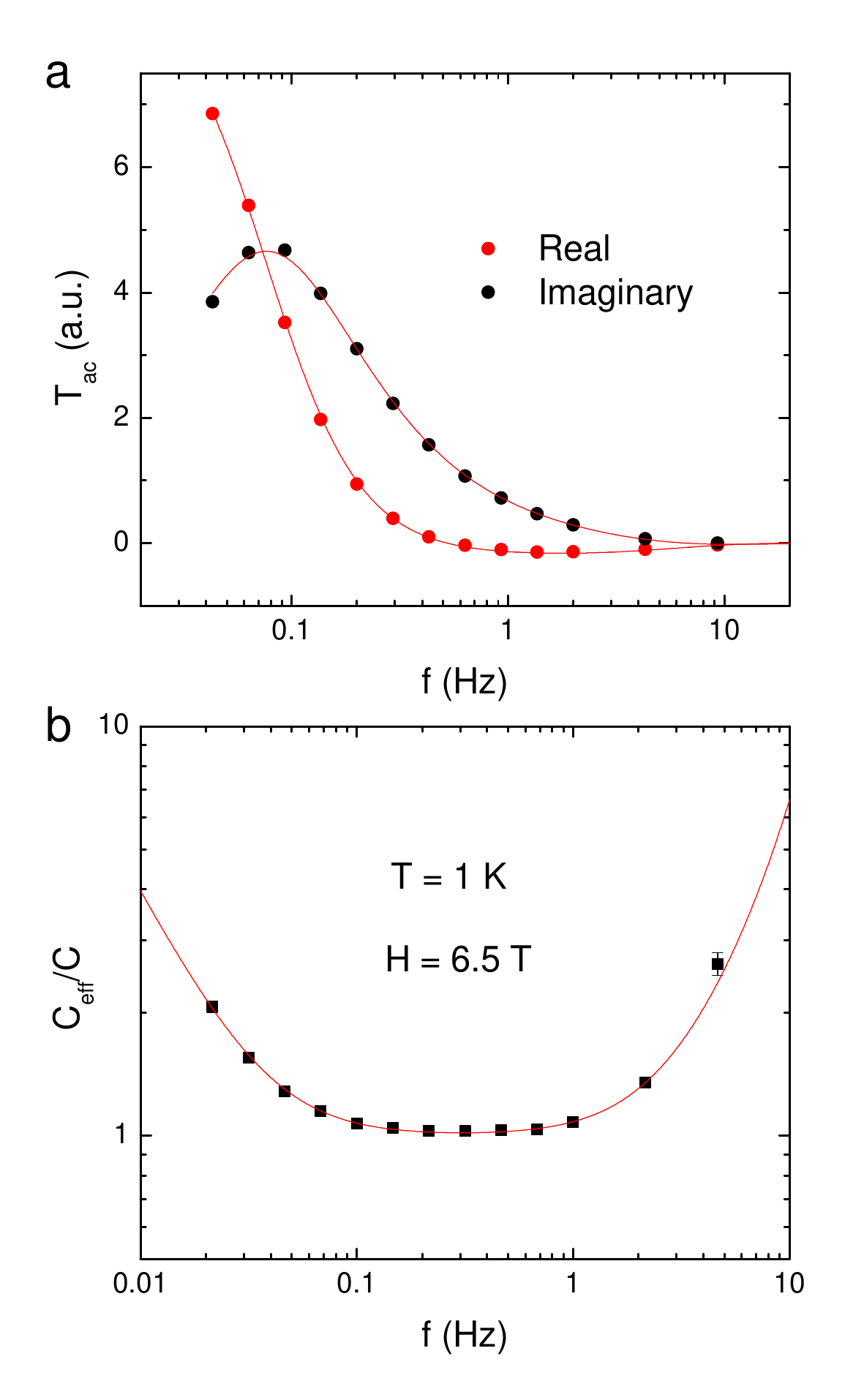}
    \caption{\label{Fit}  
   {\bf  Spectra of the observed complex temperature and the inferred effective heat capacity at 1 Kelvin and 6.5 Tesla.} Panel {\bf a} shows the frequency dependence of the real (red dots) and imaginary part (black dots) of complex temperature $\hat{T}_{ac}$. The thin red curves show the fits to Eq. \ref{eq:Tac}. (Panel {\bf b}) Frequency dependence of the effective heat capcacity (black squres) and the fit (red curve). The effective heat capacity C$_{eff}$ coincides with the sample heat capacity $C=P_0/(2\omega|\hat{T}_{ac}(\omega^*)|\,)$ when the frequency $\omega^*$ falls within the sweet spot $\omega\tau_1\ll 1 \ll \omega\tau_{ext}$ (see text). Throughout, $f = \omega/2\pi$ is the frequency of the AC power.} 
\end{figure*}

In the experiment, the measurements extended from 0.45 K to 30 K in temperature and from 0 to 8 T in field. In each representative region of 
the $T$-$H$ plane investigated, we have measured the frequency spectrum of the effective heat capacity $C_{eff}(\omega)\equiv P_0/2\omega |\hat{T}_{ac}(\omega)| $. We carried out fits to the equations above, and found the optimal frequency $\omega^*$ in each region of the $T$-$H$ plane. The fit to one of the spectra is shown in Figure~\ref{Fit}. The flat portion of the spectrum corresponds to the sweet spot in which $C$ is identified with $P_0/2\omega |\hat{T}_{ac}|$. The fits to the real and imaginary parts of Eq. \ref{eq:Tac} 
are also shown in the inset of Figure~\ref{Fit}. 

\begin{figure*}
    \includegraphics[width=14cm]{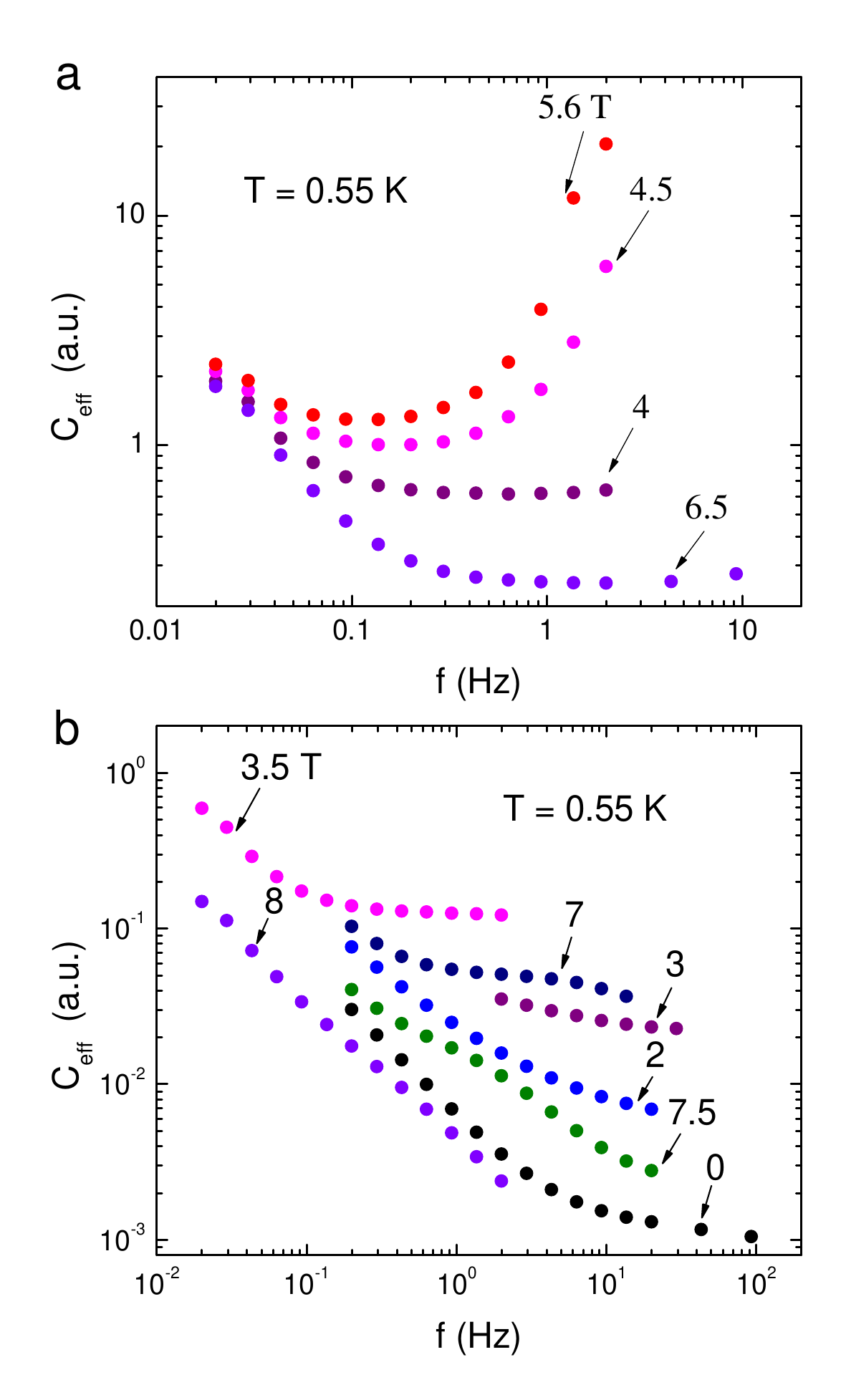}
    \caption{\label{0.55K} 
  {\bf   Spectra of the effective heat capacity $C_{eff}$ at 0.55 K at selected fields.} ({\bf a}) Frequency dependence of the sweet spot near the quantum critical point. As the magnetic field reaches the quantum critical point, the sweet spot moves to lower frequencies, reflecting the increase of the heat capacity. ({\bf b}) Glassy behavior below 3.5 T and above 7 T. Effective heat capacity C$_{eff}$ decreases monotonically as the frequency increases. No sweet spot was founded in this field range. For clarity, the curves have been shifted vertically to avoid overlap.} 
\end{figure*}

Figure \ref{0.55K} (upper panel) displays the spectrum of $C_{eff}$ at 0.55 K at several values of $H$. The evolution of the sweet spot as $H$ varies is apparent. As $H$ approaches the critical field $H_\mathrm{c}$, the sweet spot moves to lower frequencies, reflecting the increase of the heat capacity of the sample.

An important benefit of mapping the spectra over the entire $T$-$H$ plane is that we can 
observe the onset of glassy behavior. In the glass-like regions (which appear below 1 K in specific field ranges), 
the spectra decrease monotonically with increasing $\omega$.
The flat portion is not observed. Several traces for $H$
below 3.5 T and above 7 T are shown in the lower panel of Fig. \ref{0.55K} (all curves are at 0.55 K). 
In these regimes (demarcated by the dashed curves in 
Fig. \ref{figphase}), the measured spectrum cannot be fitted to Eq.~\ref{eq:Tac}, 
so we cannot extract a value for $C$.

\begin{figure*}
    \includegraphics[width=14cm]{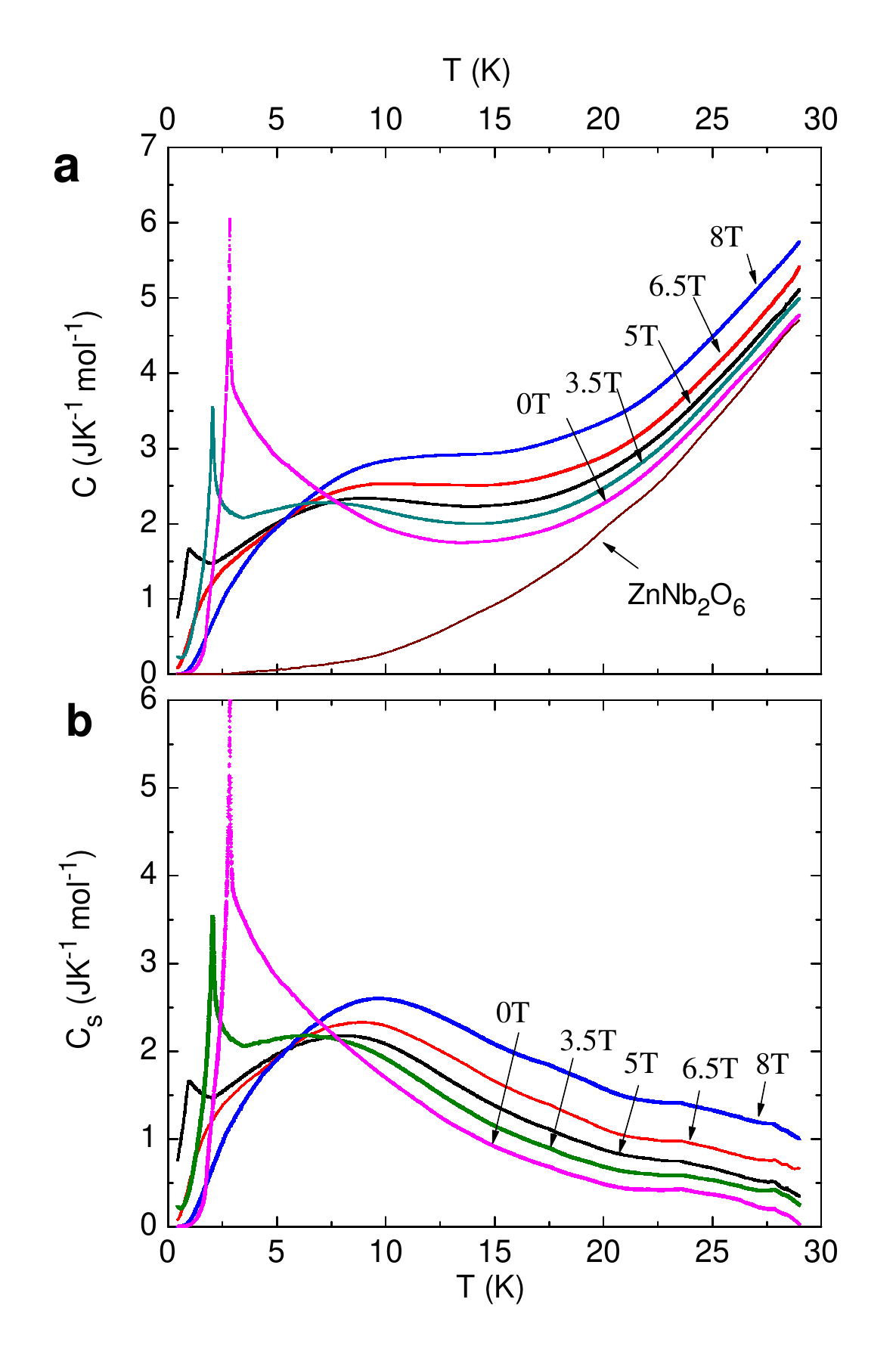}
    \caption{\label{Phonon} 
{\bf		Behavior of the heat capacity measured up to 30 K at selected magnetic fields.} ({\bf a}) The bold curves are the total heat capacity $C$ measured below 30 K in fixed field $H$ = 0--8 T. The thin curve (wine colored) shows the heat capacity of the nonmagnetic analog ZnNb$_2$O$_6$ measured by Hanawa \etal. We slightly rescaled their data by $\sim$ 13\% to match our measured curves (the disagreement arises from uncertainties in estimating the crystal size). The phonon contribution to the total heat capacity is negligible below 4 K. Panel ({\bf b}) plots the spin heat capacity $C_s$ obtained after subtraction of the phonon contribution at the selected $H$.
    }
\end{figure*}

\vspace{0.3cm}\noindent{\bf Phonon Contribution}
Below 4 K, the contribution of the phonon to the heat capacity is negligible. 
However, above 4 K, the phonon contribution to the observed $C$ is substantial.
To isolate the heat capacity of the spin degrees of freedom, we 
estimate the phonon contribution in CoNb$_2$O$_6$ as equivalent to the heat capacity in ZnNb$_2$O$_6$ measured 
by Hanawa \etal ~\cite{Hanawa}.
ZnNb$_2$O$_6$ is a nonmagnetic analog of CoNb$_2$O$_6$ with nearly identical lattice structure. 
In the upper panel of Fig~\ref{Phonon}, the raw curves for CoNb$_2$O$_6$ (before the phonon subtraction) are plotted.
The heat capacity of ZnNb$_2$O$_6$ is also plotted alongside with a slight rescaling 
(by 13 $\%$) to achieve asymptotic agreement with our zero-$H$ curve when $T$ exceeds 25 K. 
The rescaling is consistent with the combined experimental uncertainties in the 2 experiments. 
Curves of the spin contribution to the heat capacity $C_s(T)$, obtained after phonon subtraction, are plotted 
in the lower panel of Fig~\ref{Phonon}.

%%%%%%%%%%%%%%%%%%%%%%%%%%%%%%%%%%%%%%%%%%%

\vspace{0.3cm}\noindent{\bf Free-fermion solution}
We use the free-fermion solution~\cite{Lieb,Pfeuty} of the 1D Transverse Ising Model to calculate the heat capacity of the spin d.o.f.
The TIM Hamiltonian is
\be
H  = - \Gamma\sum_i S^z_i - J\sum_i S^x_iS^x_{i+1},
\label{eq:H}
\ee
with $\Gamma$ the applied transverse field (along $\bf\hat{z}$) and $J$ the easy-axis exchange (along $\bf\hat{x}$). 
The spin operators $S^x_i$ and $S^y_i$, expressed in the combination 
\be
a_i^{\dagger} = S^x_i + iS^y_i; \quad a_i = S^x_i - iS^y_i,
\ee
are converted by the Jordan-Wigner transformation into the fermion operators
\be
c_i = \exp(\pi i\sum^{i-1}_{j=1} a_j^{\dagger}a_j)a_i, \quad c_i^{\dagger} = a_i^{\dagger} \exp(-\pi i\sum^{i-1}_{j=1} a_j^{\dagger} a_j).
\label{eq:c}
\ee
$H$ is then reduced to terms bilinear in the fermion operators. A final Bogolyubov transformation to the new fermion operators 
$\eta^\dagger_k$, $\eta_k$ achieves diagnonalization, viz.
\be
H = \Gamma \sum_k\Lambda_k \eta_k^\dagger \eta_k -\frac{\Gamma}{2}\sum_k \Lambda_k.
\label{eq:Heta}
\ee
The free-fermion excitation energy is $\varepsilon_k = \Gamma\Lambda_k$, with 
\be
\Lambda_k = \sqrt{1+\lambda^2 + 2\lambda\cos k}; \quad \lambda = J/2\Gamma.
\label{eq:Lambda}
\ee
The free energy is given by
\be
F = -Nk_BT\left[ \ln 2 + \int^\pi_0 \frac{dk}{\pi} \ln\cosh(\frac12\beta\Gamma\Lambda_k) \right],
\label{eq:F}
\ee
with $\beta = 1/(k_BT)$. From $F$, we obtain the molar heat capacity in Eq. \ref{eq:Cs}.

\noindent\vspace{3mm}\\

{\bf Acknowledgements:} We acknowledge valuable discussions with N.P. Armitage, Z. C. Gu, F.D.M. Haldane and D. A. Huse. 
Research at Princeton was supported by an NSF-MRSEC grant (DMR 1420541).
Crystal growth and materials synthesis work at IQM was supported by U.S. 
Department of Energy, Office of Basic Energy Sciences, under award DE-FG02-08ER46544.
T.L acknowledges a scholarship from Japan Student Services Organization.

\vspace{3mm}
{\bf Author contributions} \\\noindent
T. L. designed the experiment and performed the measurements. 
S. M. K., J. W. K., R.J. C. and T. M. M. grew the crystals and characterized their properties. 
N.P.O. and T.L. analyzed the results and wrote most of the main text.
All authors shared ideas and carefully read the manuscript.

\vspace{3mm}
{\bf Additional information}\\\noindent
Competing financial interests: The authors declare no competing financial interests.

\newpage
%%%%%%%%%%%%%%%%%%%%%%%%%%%%%%%%%%
%%%%%%%%%%%%%%%%%%%%%%%%%%%%%%%%%%
%%%%%%%%%%%%%%%%%%%%%%%%%%%%%%%%%% FIGURE 1

%%%%%%%%%%%%%%%%%%%%%%%%%%%%%%%%%%
%%%%%%%%%%%%%%%%%%%%%%%%%%%%%%%%%%
%%%%%%%%%%%%%%%%%%%%%%%%%%%%%%%%%% FIGURE 2

%

%%%%%%%%%%%%%%%%%%%%%%%%%%%%%%%%%%
%%%%%%%%%%%%%%%%%%%%%%%%%%%%%%%%%%
%%%%%%%%%%%%%%%%%%%%%%%%%%%%%%%%%% FIGURE 3

%%%%%%%%%%%%%%%%%%%%%%%%%%%%%%%%%%
%%%%%%%%%%%%%%%%%%%%%%%%%%%%%%%%%%
%%%%%%%%%%%%%%%%%%%%%%%%%%%%%%%%%% FIGURE 4

%
%
%%%%%%%%%%%%%%%%%%%%%%%%%%%%%%%%%%
%%%%%%%%%%%%%%%%%%%%%%%%%%%%%%%%%%
%%%%%%%%%%%%%%%%%%%%%%%%%%%%%%%%%% FIGURE 5

%
%
%%%%%%%%%%%%%%%%%%%%%%%%%%%%%%%%%%
%%%%%%%%%%%%%%%%%%%%%%%%%%%%%%%%%%
%%%%%%%%%%%%%%%%%%%%%%%%%%%%%%%%%% FIGURE 6

%
%
%%%%%%%%%%%%%%%%%%%%%%%%%%%%%%%%%%
%%%%%%%%%%%%%%%%%%%%%%%%%%%%%%%%%%
%%%%%%%%%%%%%%%%%%%%%%%%%%%%%%%%%% FIGURE 7

%
%%%%%%%%%%%%%%%%%%%%%%%%%%%%%%%%%%
%%%%%%%%%%%%%%%%%%%%%%%%%%%%%%%%%%
%%%%%%%%%%%%%%%%%%%%%%%%%%%%%%%%%% FIGURE 8

\begin{thebibliography}{99}

\bibitem{Sachdev} Sachdev, S. 
Quantum Phase Transition (Cambridge Univ. Press, 1999).

\bibitem{Aeppli} Ghosh, S., Rosenbaum, T.F., Aeppli, G. $\&$ Coppersmith, S.N., 
Entangled quantum state of magnetic dipoles.
Nature {\bf 425}, 48-51 (2003).


\bibitem{Coldea} Coldea, R. \etal,
%D. A. Tennant, E. M. Wheeler, E. Wawrzynska, D. Prabhakaran, M. Telling, K. Habicht, P. Smeibid, $\&$ K. Kiefer,
Quantum criticality in an Ising chain: experimental evidence for emergent E8 symmetry. 
Science {\bf 327}, 177-180 (2010).



\bibitem{E8} Zamolodchikov, A. B.,
Integrals of motion and S-matrix of the (scaled) $T\,=\,T_\mathrm{c}$ Ising model with magnetic field. 
Int. J. Mod. Phys. A {\bf 4}, 4235-4248 (1989).


\bibitem{Delfino} Delfino, G. 
Integrable field theory and critical phenomena: the Ising model in a magnetic field.
J. Phys. A: Math. Gen. {\bf 37}, R45–R78 (2004).

\bibitem{Kitaev} Kitaev, A. Yu. 
Unpaired Majorana fermions in quantum wires.
Physics-Uspekhi {\bf 44}, 131-136  (2001), doi:10.1070/1063-7869/44/10S/S29.

\bibitem{Fendley} Fendley, P. 
Parafermionic edge zero modes in Zn-invariant spin chains.
Jnl. Statistical Mechanics {\bf 1211}, P11020 (2012),  DOI: 10.1088/1742-5468/2012/11/P11020 (2012)

\bibitem{Alicea} Alicea, J.
New directions in the pursuit of Majorana fermions in solid state systems.
Rep. Prog. Phys. {\bf 75}, 076501 (2012).


\bibitem{Cabrera} Cabrera, I. \etal, 
%J. D. Thompson, R. Coldea, D. Prabhakaran, R. I. Bewley, T. Guidi, J. A. Rodriguez-Rivera and C. Stock
Excitations in the quantum paramagnetic phase of the quasi-one-dimensional Ising magnet
CoNb$_2$O$_6$ in a transverse field: Geometric frustration and quantum renormalization effects.
Phys. Rev. B {\bf 90}, 014418 (2014).


\bibitem{Armitage} Morris, C. M. \etal,
%R. Valdés Aguilar, A. Ghosh, S. M. Koohpayeh, J. Krizan, R. J. Cava,O. Tchernyshyov, T. M. McQueen, and N. P. Armitage
Hierarchy of Bound States in the One-Dimensional Ferromagnetic Ising Chain CoNb2O6 Investigated by High-Resolution Time-Domain Terahertz Spectroscopy.
Phys. Rev. Lett. {\bf 112}, 137403 (2014).



\bibitem{Imai} Kinross, A.W. \etal,
%M. Fu, T. J. Munsie, H. A. Dabkowska, G. M. Luke, Subir Sachdev, and T. Imai,
Evolution of Quantum Fluctuations Near the Quantum Critical Point of the Transverse Field Ising Chain System CoNb$_2$O$_6$.
Phys. Rev. X {\bf 4}, 031008 (2014).


\bibitem{Hanawa} Hanawa, T., Shinkawa, K., Ishikawa, M., Miyatani, K., Saito, K. $\&$ Kohn, K. 
Anisotropic Specific Heat of CoNb$_2$O$_6$ in Magnetic Fields.
Jnl. Phys. Soc. Jpn. {\bf 63}, 2706-2715 (1994).


\bibitem{Kobayashi} Kobayashi, S., Mitsuda, S. $\&$ Prokes, K. 
Low-temperature magnetic phase transitions of the geometrically frustrated isosceles triangular Ising antiferromagnet CoNb$_2$O$_6$. 
Phys. Rev. B {\bf 63}, 024415 (2000).


\bibitem{Balents} Lee, S. B., Kaul, R. K. $\&$ Balents, L. 
Interplay of quantum criticality and geometric frustration in columbite. 
Nat. Phys. {\bf 6}, 702-706 (2010).



\bibitem{Lieb} Lieb, E. H., Schultz, T. D. $\&$ Mattis, D. C.
Two soluble models of an antiferromagnetic chain. 
Ann. Phys. {\bf 16}, 407-466 (1961).

\bibitem{Lieb2} Schultz, T. D., Mattis, D. C. $\&$ Lieb, E. H. 
Two-Dimensional Ising Model as a Soluble Problem of Many Fermions.
Rev. Mod. Phys. {\bf 36}, 856-871 (1964).

\bibitem{Pfeuty} Pfeuty, P.
The one-dimensional Ising model with a transverse field. 
Ann. Phys. {\bf 57}, 79-90 (1970).




\bibitem{Collan1}	Collan, H. K., Krusius, M. $\&$ Pickett, G. R. 
Suppression of the Nuclear Heat Capacity in Bismuth Metal by Very Slow Spin-Lattice Relaxation, and a New Value for the Electronic Specific Heat.
Phys. Rev. Lett. {\bf 23}, 11-13 (1969).

\bibitem{Collan2}	Collan, H. K., Krusius, M. $\&$ Pickett, G. R. 
Specific Heat of Antimony and Bismuth between 0.03 and 0.8 K. 
Phys. Rev. B {\bf 1}, 2888-2895 (1970).


\bibitem{McCoy} McCoy, B. M. $\&$ Wu, T. T. 
Two dimensional Ising field theory in a magnetic field: breakup of the cut in the two-point function. 
Phys. Rev. D {\bf 18}, 1259-1267 (1978).



\bibitem{Rutkevich} Rutkevich, S.B. 
Energy Spectrum of Bound-Spinons in the Quantum Ising Spin-Chain Ferromagnet. 
J Stat Phys {\bf 131}, 917–939 (2008).

\bibitem{Sullivan} Sullivan, P. F. $\&$ Seidel, G.
Steady-State, ac-Temperature Calorimetry. 
Phys. Rev. {\bf 173}, 679-685 (1968).


\end{thebibliography}
\end{document}